\documentclass[galaxies,article,accept,moreauthors,pdftex,10pt,a4paper]{mdpi}

\firstpage{1}
\articlenumber{10}
\doinum{10.3390/galaxies4030010}
\pubvolume{4}
\pubyear{2016}
\copyrightyear{2016}
\externaleditor{Academic Editor: Jose L. G\'{o}mez, Alan P. Marscher and Svetlana G. Jorstad }
\history{Received: 12 July 2016; Accepted: 18 August 2016}
\pdfoutput=1

\usepackage{subfigure}
\makeatletter
\renewcommand{\@thesubfigure}{\normalsize(\textbf{\alph{subfigure}})}
\makeatother

\usepackage{booktabs}
\usepackage{multirow}
\usepackage{soul}
\usepackage{microtype}

\newcommand\myurl[1]{\changeurlcolor{black}\url{#1}\changeurlcolor{blue}}
\makeatletter
\g@addto@macro{\UrlBreaks}{\UrlOrds}
\makeatother

\Title{Precessing Jet in the High-Redshift Blazar J0017+8135}

\Author{Krist\'{o}f Rozgonyi $^{1}$ and S\'{a}ndor {Frey} $^{2,}$*}

\AuthorNames{Krist\'{o}f Rozgonyi and S\'{a}ndor Frey }

\address{%
$^{1}$ \quad Department of Physics of Complex Systems, E\"{o}tv\"{o}s Lor\'{a}nd {University}, P\'{a}zm\'{a}ny P. s\'{e}t\'{a}ny 1/A, Budapest {H-1117}, Hungary; rstofi@gmail.com\\

$^{2}$ \quad F\"{O}MI Satellite Geodetic Observatory, P.O. Box 585, Budapest {H-1592}, Hungary}
\corres{Correspondence: frey.sandor@fomi.hu}

\abstract{The prominent flat-spectrum radio quasar J0017+8135 (S5 0014+81) at $z$ = 3.366 is one of the most luminous active galactic nuclei (AGN) known. Its milliarcsecond-scale radio jet structure has been studied with very long baseline interferometry (VLBI) since the 1980s.
The quasar was selected as one of the original defining objects of the International Celestial Reference Frame, but left out from its current second realization (ICRF2) because of systematic long-term positional variations. Here we analyse archival  8.6- and 2.3-GHz VLBI imaging data collected at nearly 100 different epochs during more than 20 years, to obtain information about the kinematics of jet components. Because of the cosmological time dilation, extensive VLBI monitoring data are essential to reveal changes in the jet structure of high-redshift AGN.
In the case of J0017+8135, the data can be described with a simple kinematic model of jet precession with a 12-year periodicity in the observer's frame.}


\begin{document}



\section{Introduction}

The blazar J0017+8135 (S5 0014+81) is listed in the S5 sample of strong northern radio sources with $S$ = 551 mJy total flux density at 5~GHz \cite{1}. Its spectroscopic redshift is $z$ = 3.366 \cite{2}. This flat-spectrum radio quasar is known as one of the most luminous active galactic nuclei (AGN) and has been subject to various studies at multiple wavebands, from the radio to X-rays (e.g., \cite{3,4,5,6}). The mass of its central black hole is $M = 4 \times 10^{10} M_{\odot}$, among the highest values measured in AGN \cite{5}. The source has been extensively studied with very long baseline interferometry (VLBI) since the 1980s, and it was also one of the sources first observed with space VLBI \cite{7,8}. The one-sided core--jet structure of J0017+8135 extending towards the south at milliarcsecond (mas) angular scales is typical for blazars. The jet components show no or moderate apparent superluminal motion \cite{9,10}. J0017+8135 was selected as one of the original defining objects of the International Celestial Reference Frame (ICRF), but left out from its current second realization (ICRF2) because of systematic long-term positional variations~\cite{11}. In this paper, we make use of the publicly available archival 2.3- and 8.6-GHz VLBI imaging data spanning more than two decades, to investigate the positional variations of the jet components in J0017+8135. We find indication of periodic changes and attempt to explain them in the framework of a simple jet precession model. A detailed jet kinematic study for a blazar at such a high redshift is unique. Changes in the source structure appear slower by a factor of $(1+z)$ in the observer's frame due to cosmological time dilation; therefore, long-term VLBI monitoring is necessary to reliably reveal them in the case of high-redshift radio-loud AGN.

\section{VLBI Observations, Images and Brightness Distribution Modeling}

The study presented here is based on 95 VLBI imaging data sets at 8.6 GHz (X band). In 74~cases, VLBI data obtained at the same epochs are also available at 2.3 GHz (S band). The observations were done with the U.S. National Radio Astronomy Observatory (NRAO) Very Long Baseline Array (VLBA), occasionally supplemented with various other radio telescopes, covering the time range from August~1994 to February 2015. We used the calibrated visibility data sets downloaded from \cite{12}.

\begin{figure}[H]
   \centering
	 \subfigure[]{\includegraphics[width=0.47\textwidth]{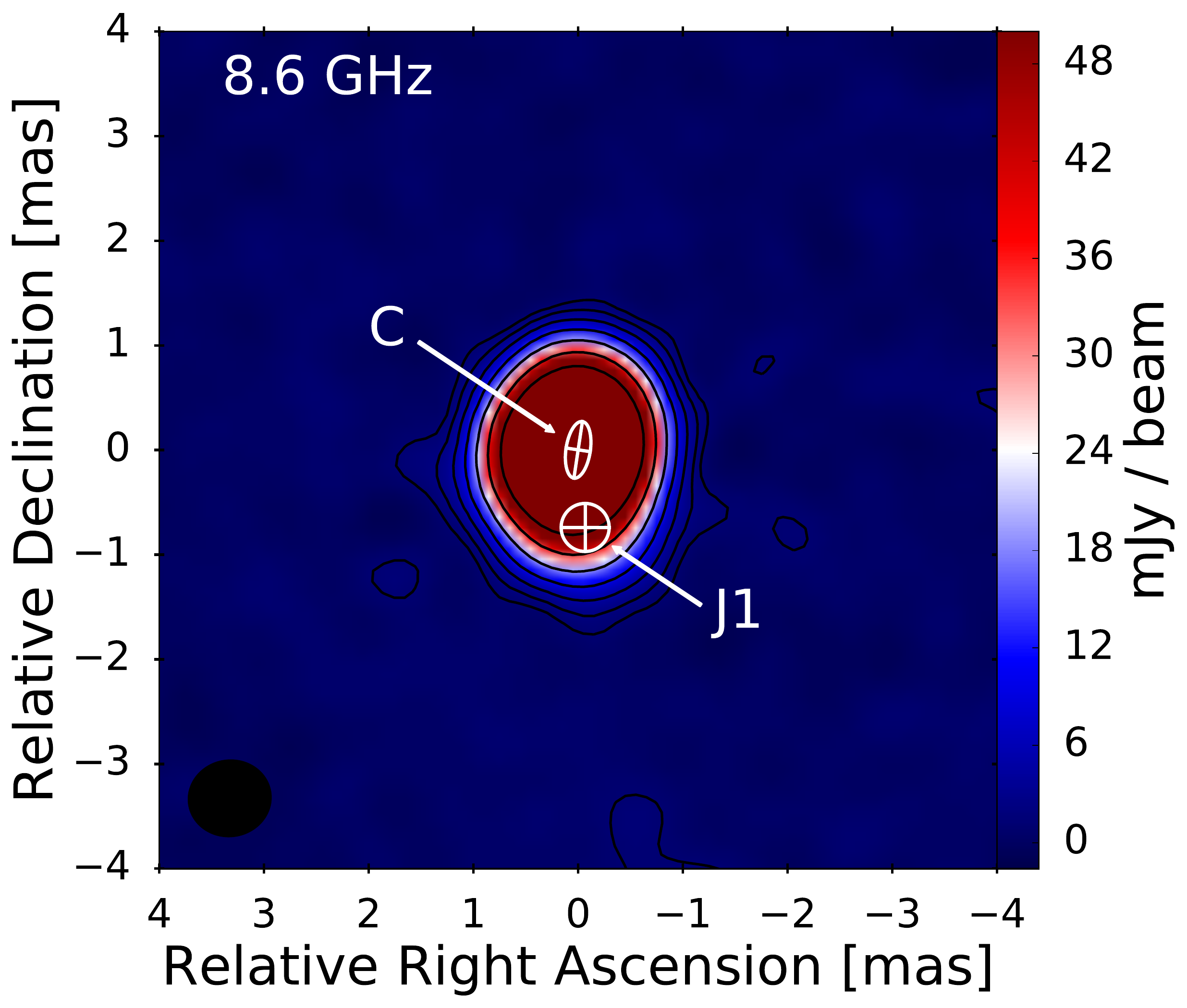}}
	\subfigure[]{ \includegraphics[width=0.47\textwidth]{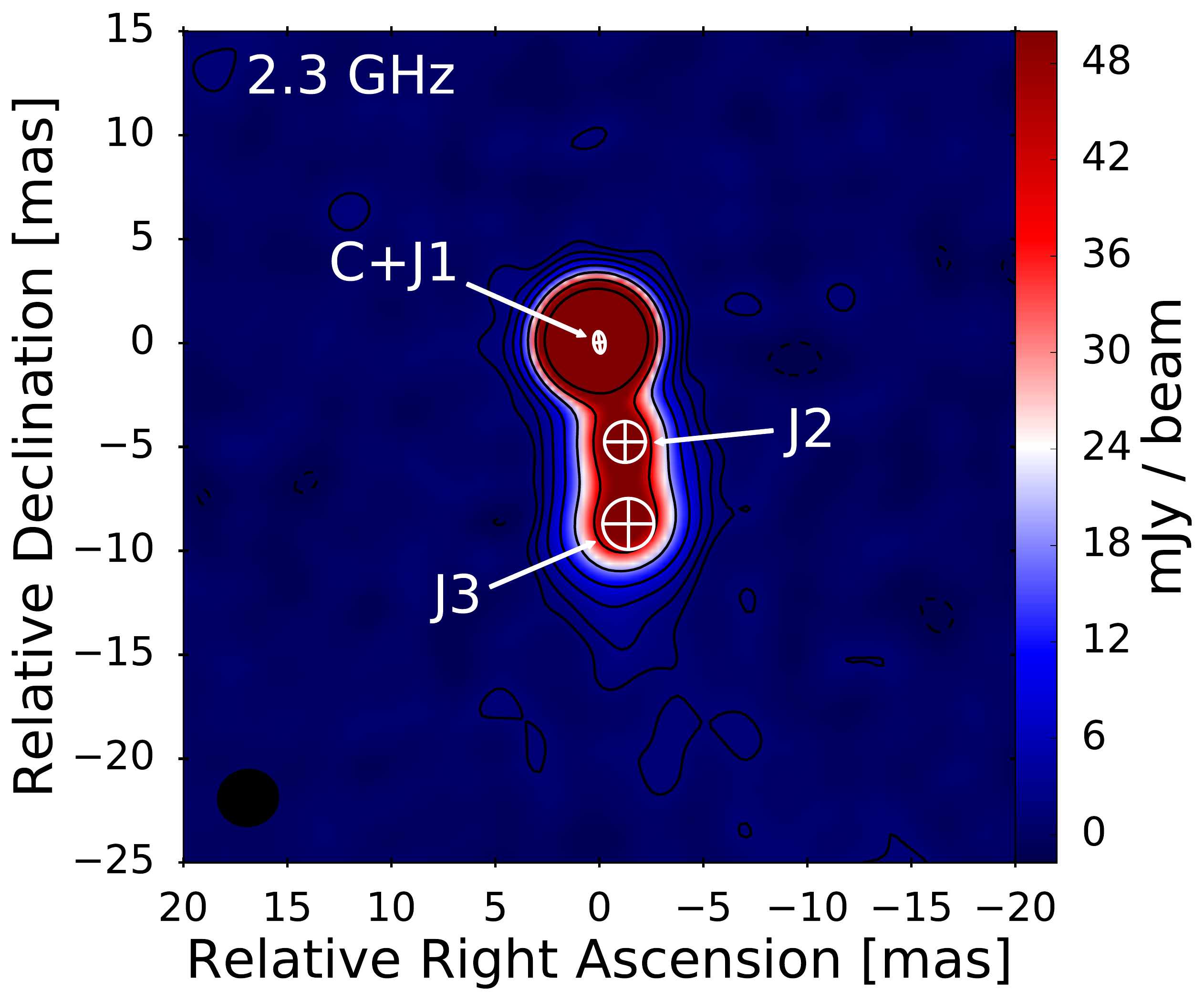}}
\caption{Pair of typical very long baseline interferometry (VLBI) images of the source J0017+8135. The images were taken in October 1998 at 8.6 GHz (\textbf{a}) and 2.3 GHz (\textbf{b}). The lowest contours correspond to 1\% of the peak brightness, the positive contours increase by a factor of 2. The white circles are centered at the positions of the fitted Gaussian model components. Their radius corresponds to the full-width-at-half-maximum diameter of the given component. The restoring beam is indicated in the lower-left corner.}
\label{images}
\end{figure}

The imaging and model-fitting were performed through standard procedures using Difmap~ \cite{13} in an automated way. The visibility data were modeled with sets of 2 or 3 Gaussian brightness distributions. The core was fitted with an elliptical Gaussian component at both frequencies. The jet at 8.6~GHz was fitted with one, and at 2.3~GHz with two circular Gaussians. A pair of typical VLBI images of the source with the fitted model components is presented in Figure~\ref{images}.

The core is labeled as C, the jet components are J1, J2 and J3, respectively. In the 2.3-GHz images, the central component is labeled as C+J1 as it blends the two innermost components found in the higher-resolution 8.6-GHz data. Although an opacity correction would be needed, it would have a minor effect due to the relatively close observing frequencies. Thus, in both cases, the central component was considered as the kinematic centre for the modeling.

\section{A Simple Jet Precession Model}

We fitted a kinematic model \cite{14} to the relative positions of the brightness distribution components measured as a function of time, to quantitatively describe the periodic behaviour of their motion. We~assumed that the plasma blobs ejected at different epochs move in a ballistic trajectory. At each epoch, we see the ejected components at their actual projected positions, and the base of the jet as the optically thick `core' region. The relative positions of the identified components can change over time if the core is wobbling (i.e., changing its absolute position). Thus, if we choose the core component as the kinematic centre of the jet, the movements of the core will be reflected in the relative movements of the ejected components. The relative positions of the jet components show periodic difference from the ballistic track, thus we assume a possible precession of the jet.

\begin{figure}[H]
\centering
\includegraphics[width=0.5\textwidth]{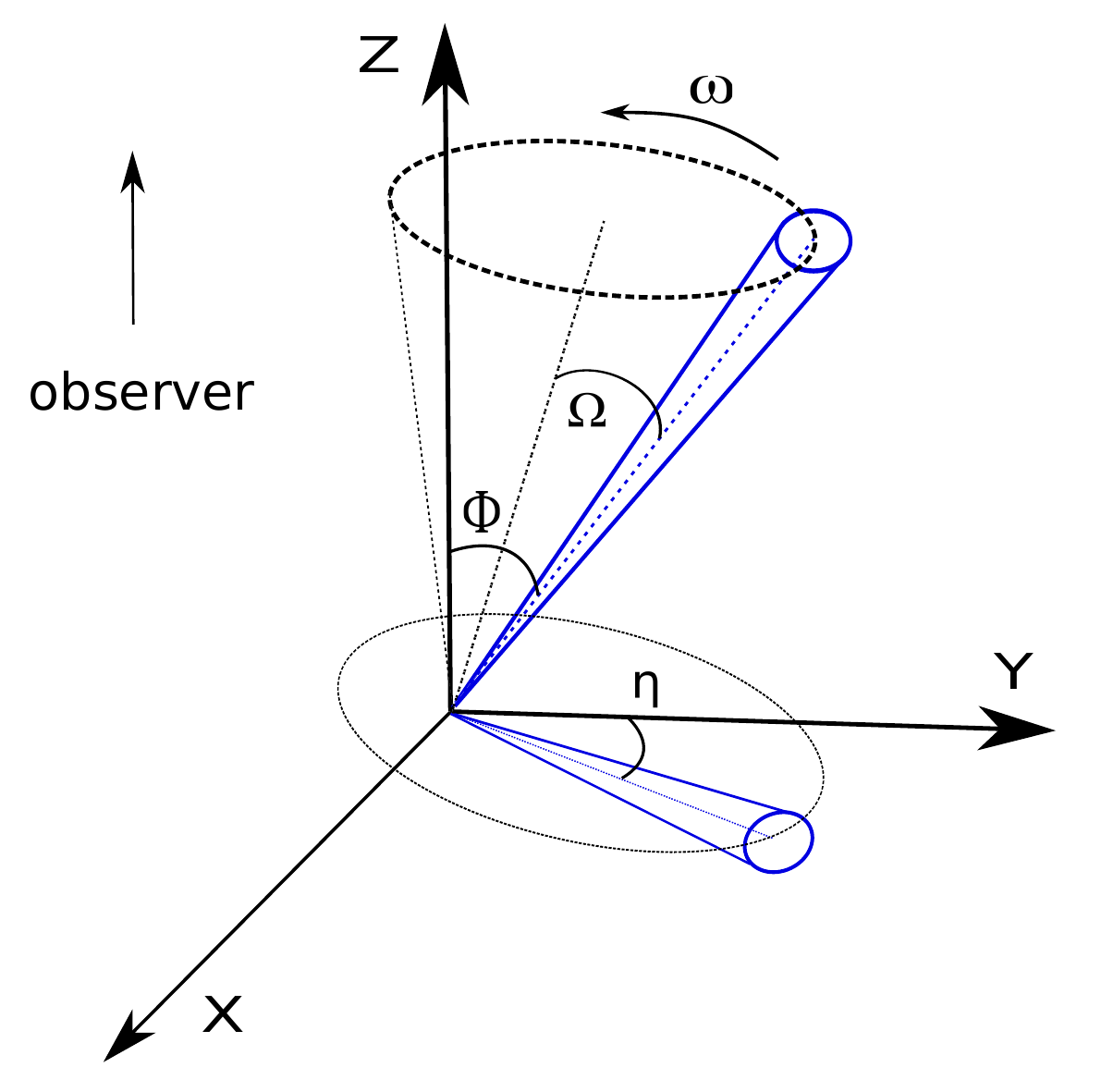}
\caption{Jet precession geometry. The line of sight is parallel to the $z$ axis. Right ascension and declination correspond to the $x$ and $y$ axes, respectively. The $\eta$ position angle is measured from north through east.}
\label{model}
\end{figure}

In our precession model, we make three assumptions for the kinematics and geometry of the jet:  (1) the jet sweeps a conical surface whose half opening angle $\Omega$ and axis are constant over time;  (2) the jet is precessing with a constant $\omega$ angular velocity around the cone axis; and  (3) the jet components ejected at $t_0$ epoch can be described at $t$ with two time-dependent angles: $\eta(t)$ and $\Phi(t)$.
In this scenario, the ejected components will follow a helical path in the plane of the sky. The projections of this kind of motion to the right ascension (RA) and declination (Dec) axes can be described as superpositions of a linear and a periodic function. The angular frequency of the projected motions is characteristic to the jet precession, thus it has to be equal in both directions. The apparent precessing motion of the jet is the combination of the two perpendicular oscillations in this model. Figure~\ref{model} shows the geometric configuration of the jet model.

Assuming this configuration, for each jet feature, we can fit a superposition of a linear and a periodic function to the projected component RA and Dec angular distances from the core:
\begin{equation}
s_i(t) = s_{0i} + \mu_it + A_i(\omega t + \varphi_i),
\end{equation}
where $ i \in [{\rm RA,Dec}] $, $s_i(t)$ is the component distance from the core at the time $t$, $s_{0i}$ is the component distance from the core at $t = 0$, $\mu_i$ is the proper motion of the component, $ \omega $ is the common angular frequency, $A_i$ and $\varphi_i$ are the fitted amplitude and phase, respectively.

As a boundary condition, the model requires that the relative RA and Dec positions of the components at the ejection epoch $t_0$ are
\begin{equation}
s_{\rm RA}(t_0) \: = \: s_{\rm Dec}(t_0) \: = \: 0.
\end{equation}

However, if the amplitude of the precession is small, we can simplify this condition so that only the fitted ballistic slopes have to intersect the $t$ axis at the same point. From this assumption, we get
\begin{equation}
  \frac{s_{0\,\rm RA}}{\mu_{\rm RA}} = \frac{s_{0\,\rm Dec}}{\mu_{\rm Dec}}.
\end{equation}

Thus we have to fit the model to the relative RA and Dec position data for all components at the same time, with the common parameter $\omega$. Considering the simplified boundary condition, the kinematic model can be expressed as
\begin{equation}
\begin{aligned}
  s_{\rm RA}(t) = \frac{s_{0\,\rm Dec}\mu_{\rm RA}}{\mu_{\rm Dec}} + \mu_{\rm RA}t + A_{\rm RA}(\omega t + \varphi_{\rm RA}), \\
  s_{\rm Dec}(t) = s_{0\,\rm Dec} + \mu_{\rm Dec}t + A_{\rm Dec}(\omega t + \varphi_{\rm Dec}).
\label{prec-model}
\end{aligned}
\end{equation}

\section{Jet Kinematic Model Parameter Estimates}

\begin{figure}[H]
\centering
\includegraphics[width=0.95\textwidth]{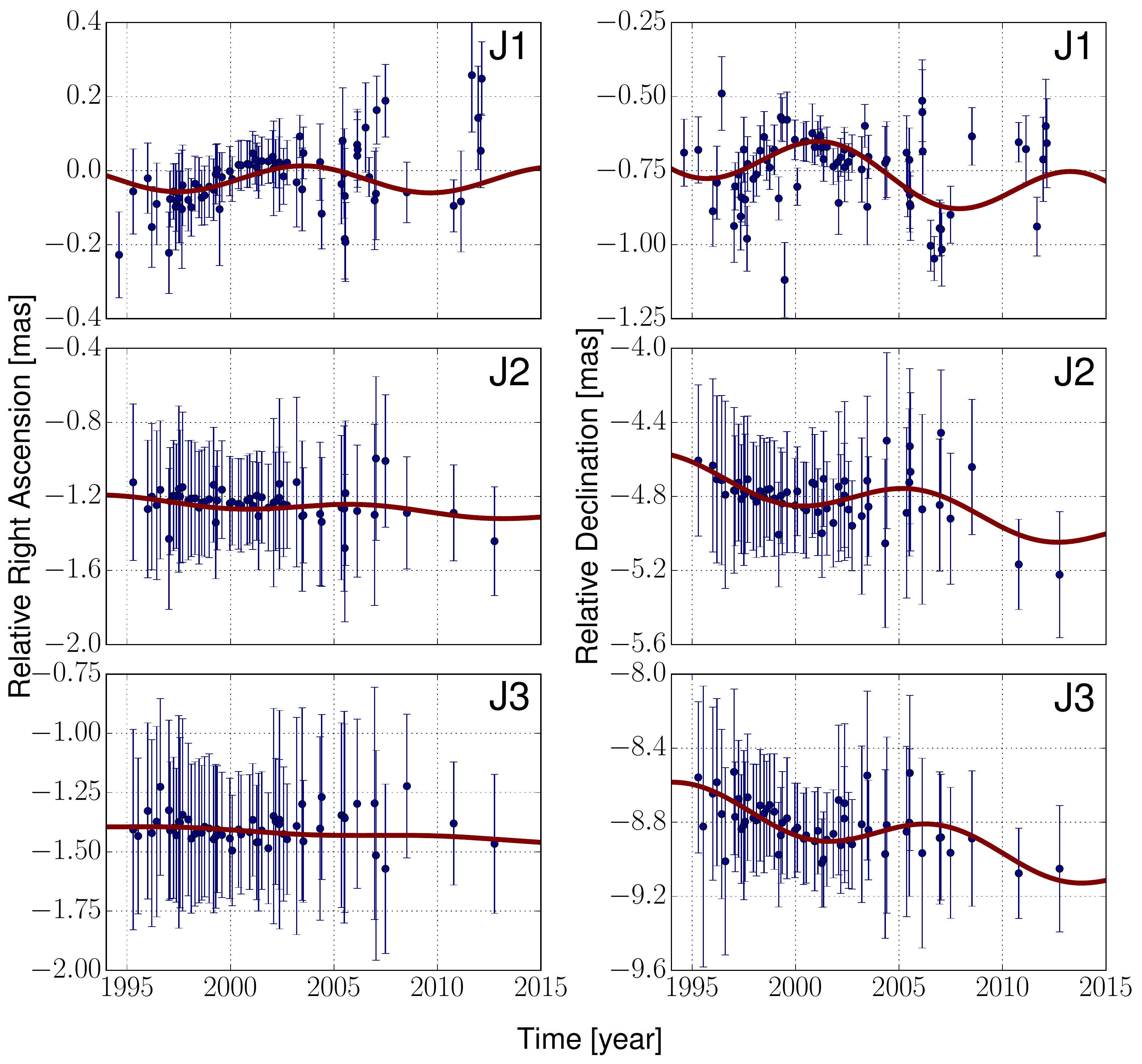}
\caption{The relative separation of the components from the core as a function of time, and the best-fitting model (solid curves). Not all components were detected at each epoch, and the dense time sampling allowed us to drop out positional outliers over 3$\sigma$ from the analysis. As a result, the fitted precession model is based on 73, 56, and 55 observing epochs for components J1, J2, and J3, respectively.}
\label{fits}
\end{figure}

To analyze the precession of the jet model components, we fitted Equation~(\ref{prec-model}) to the jet component positions relative to the core as derived from the multi-epoch VLBI data. We applied the Metropolis--Hastings algorithm \cite{15,16} to look for the best-fit parameters for each component. Figure~\ref{fits}~illustrates the fitted model for the jet components.

Based on the best-fitting model parameters, we can express the main kinematic properties of the jet components (Table~\ref{table}). We estimated the angular frequency $ \omega $, the precession period $ T $, and the apparent proper motions $ \beta_{\rm app} $ for each component (expressed in the units of the speed of light~$c$). To~estimate $ \phi(t_0) $ and $ \Omega $ with reasonably small error, we would have to observe more model components in the jet~\cite{14}. Nevertheless, we also estimated the projected angle of the precession cone axis on the plane of the sky at the ejection epoch, $ \eta(t_0) $, measured from north through east.

\begin{table}[H]
\caption{The kinematic properties of model components.}
\centering
\begin{tabular}{cccc}
\toprule
\textbf{Parameter} & \textbf{J1} & \textbf{J2} & \textbf{J3} \\
\midrule
$ \omega $ (year$^{-1}$) & $ 0.51 \: \pm \: 0.08 $ & $ 0.51 \: \pm \: 0.15 $ & $ 0.5 \: \pm \: 0.1 $ \\
$ T $ (year) & $ 12 \: \pm \: 2 $ & $ 12 \: \pm \: 4 $ & $ 12 \: \pm \: 2 $ \\
$ \beta_{\rm app} $ & $ 0.9 \: \pm \: 0.2 $ & $ 1.8 \: \pm \: 1.0 $ & $ 2.0 \: \pm \: 0.3 $ \\
$ \eta(t_0) $ ($ ^{\circ} $) & $ 180 \: \pm \: 15 $ & $ 195 \: \pm \: 34 $ & $ 189 \: \pm \: 10 $ \\
\bottomrule
\end{tabular}
\label{table}
\end{table}

\section{Discussion}

We found that, according to the expectations, the ejection epochs were in chronological order for the J3, J2, and J1 components, respectively. The apparent motion of the components is dominantly in the Dec direction, and the amplitude of the periodic motion is higher in this direction. Thus~the precession cone is projected onto the sky as an ellipse stretched in the north--south direction, in accordance with the absolute astrometric position changes \cite{11}.
Somewhat similarly to earlier results based on VLBI observations spanning shorter intervals \cite{9,10}, our analysis reveals mild superluminal motion ($\sim$$1-2c$) in the jet components of J0017+8135 but with smaller uncertainties.

In the framework of the kinematic model applied here, the jet has a precession period of ($ 12 \: \pm \: 3 $)~year in the observer's frame. This is a unique result for a blazar at $z>3$ where the cosmological time dilation makes it more difficult to detect periodicities compared to low-redshift~sources. However,~it~should be noted that this analysis is based on data that span an interval shorter than twice the putative period. Therefore any stochastic variability might be mistakenly identified as periodic. Continuing~VLBI monitoring of J0017+8135 on a longer term, and the analysis of absolute astrometric position variations \cite{11} and total flux density monitoring data \cite{17} in the context of the jet precession model would help confirming our results in the future.

Finally, it is also possible that physical mechanisms other than precession cause the observed apparently helical path of the jet components. For example, plasma instabilities propagating along a conical jet can cause this effect. In fact, studies of large samples of sources monitored with VLBI revealed that non-ballistic motion is ubiquitous in pc-scale AGN jets (\cite {18} and references therein).





\acknowledgments{This work was supported by the Hungarian National Research, Development and Innovation Office (OTKA K104539). The NRAO is a facility of the National Science Foundation operated under cooperative agreement by Associated Universities, Inc. The Astrogeo Center Database of brightness distributions, correlated flux densities, and images of compact radio sources produced with VLBI is maintained by L. Petrov.}

\authorcontributions{Krist\'{o}f Rozgonyi analyzed the data and performed the modeling, S\'{a}ndor Frey contributed to the design of the study and the interpretation of the results. Both authors participated in the writing of the paper.}

\conflictofinterests{The authors declare no conflict of interest.}





\bibliographystyle{mdpi}

\renewcommand\bibname{References}



\end{document}